\documentclass[%
reprint,
superscriptaddress,
amsmath,amssymb,
aps,
]{revtex4-1}

\usepackage{graphicx}% Include figure files
\usepackage{dcolumn}% Align table columns on decimal point
\usepackage{bm}% bold math

\usepackage{todonotes} % Comments

\begin{document}
	
	\preprint{APS/123-QED}
	
	\title{Three dimensional waveguide-interconnects for scalable integration\\ of photonic neural networks}
	
	\author{Johnny Moughames}
	\affiliation{Institut FEMTO-ST,  Universit\'e Bourgogne Franche-Comt\'e CNRS UMR 6174, Besan\c{c}on, France}
	
	\author{Xavier Porte}
	\altaffiliation[]{Corresponding author: javier.porte@femto-st.fr}
	\affiliation{Institut FEMTO-ST,  Universit\'e Bourgogne Franche-Comt\'e CNRS UMR 6174, Besan\c{c}on, France}
	
	\author{Michael Thiel}
	\affiliation{Nanoscribe GmbH, Hermann-von-Helmholtz-Platz 6, 76344 Eggenstein-Leopoldshafen, Germany}
	
	\author{Gwenn Ulliac}
	\affiliation{Institut FEMTO-ST,  Universit\'e Bourgogne Franche-Comt\'e CNRS UMR 6174, Besan\c{c}on, France}
	
	\author{Maxime Jacquot}
	\affiliation{Institut FEMTO-ST,  Universit\'e Bourgogne Franche-Comt\'e CNRS UMR 6174, Besan\c{c}on, France}
	
	\author{Laurent Larger}
	\affiliation{Institut FEMTO-ST,  Universit\'e Bourgogne Franche-Comt\'e CNRS UMR 6174, Besan\c{c}on, France}
	
	\author{Muamer Kadic}
	\affiliation{Institut FEMTO-ST,  Universit\'e Bourgogne Franche-Comt\'e CNRS UMR 6174, Besan\c{c}on, France}
	
	\author{Daniel Brunner}
	\affiliation{Institut FEMTO-ST,  Universit\'e Bourgogne Franche-Comt\'e CNRS UMR 6174, Besan\c{c}on, France}

\date{\today}% It is always \today, today,

\begin{abstract}
Photonic waveguides are prime candidates for integrated and parallel photonic interconnects.
Such interconnects correspond to large-scale vector matrix products, which are at the heart of neural network computation.
However, parallel interconnect circuits realized in two dimensions, for example by lithography, are strongly limited in size due to disadvantageous scaling.
We use three dimensional (3D) printed photonic waveguides to overcome this limitation.
3D optical-couplers with fractal topology efficiently connect large numbers of input and output channels, and we show that the substrate's footprint area scales linearly.
Going beyond simple couplers, we introduce functional circuits for discrete spatial filters identical to those used in deep convolutional neural networks.
\end{abstract}

\maketitle
	
\section{Introduction}
	
The interconnection of numerous input and output channels (IO-channels) is the basic operation behind many applications.
A parallel and energy-efficient interconnect has therefore been a desired technology for decades \cite{Shamir1989,Lee1989}, finding use in diverse fields such as telecommunication, inter and intra-chip data buses and potentially endoscopy \cite{Choudhury2019}.
Most timely, it also is highly desired for connecting layers of deep neural networks to efficiently provide the typically large scale vector-matrix products \cite{LeCun2015}.

The integration of such an apparatus is challenging.
To achieve parallelism, serial routing is naturally not an option, and a large number of direct physical links connecting the IO-channels is required.
Such channel multiplexing can be created in different dimensions like wavelength or space, and here we address spatial multiplexing.
If a direct connection architecture is realized electronically, the strong capacitive interactions between long connection wires will result in prohibitive energy dissipation and bandwidth limitations \cite{Miller2017,Esmaeilzadeh2012}.
There are additional, more practical challenges.
Lithographic fabrication typically integrates circuits in two dimensions (2D), and a 2D interconnect's footprint grows quadratic with the number of IO-channels.
The cross-bar interconnect illustrates this fundamental relationship.

Optical routing removes the energy dissipation associated to charging the capacity of electronic signaling wires \cite{Miller2017}, and free-space interconnects with many IO-channels have long been explored \cite{Shamir1989,Lee1989}.
Integrated photonic interconnects, however, remain size-limited by the unfavourable scaling between area and the number of IO-channels in 2D \cite{Miller2015, Shen2016, Hughes2018, Peng2018}.
Crucially, the same scaling is found for wavelength division multiplexing.

We demonstrate the integration of such photonic interconnects in 3D for the first time.
Complex 3D-routed waveguides are created by two photon polymerization \cite{Deubel2004,Yang2019}.
We introduce a fractal architecture which efficiently connects many IO-channels, and we demonstrate an integrated photonic interconnect of unreported size hosting 225 input and 529 output channels within a footprint area of only 0.46$\times$0.46~mm$^2$.
Crucially, this footprint area scales linearly.
Such a printed photonic circuit can fully and in parallel connect the layers of deep neural networks of a commercially relevant size \cite{LeCun2015,Jouppi2017}.
Going beyond, we demonstrate a 3D-waveguide architecture implementing 9 spatial filters with a Haar convolution Kernel \cite{NIPS2006_3130} of stride and width 3.
Such convolutional filters represent a fundamental operation of deep convolutional neural networks \cite{LeCun2015}.

3D photonic circuits promise integrated, parallel, scalable and hence large interconnects with potentially low energy dissipation.
Our concept is based on mature fabrication technology which has also been exploited for photonic wirebonding between chips \cite{Lindenmann2012,Koos2013}.

\section{Scaling of interconnects}
\label{sec:3DTopology}

\begin{figure}[t]
	\centering
	\includegraphics[width=\linewidth]{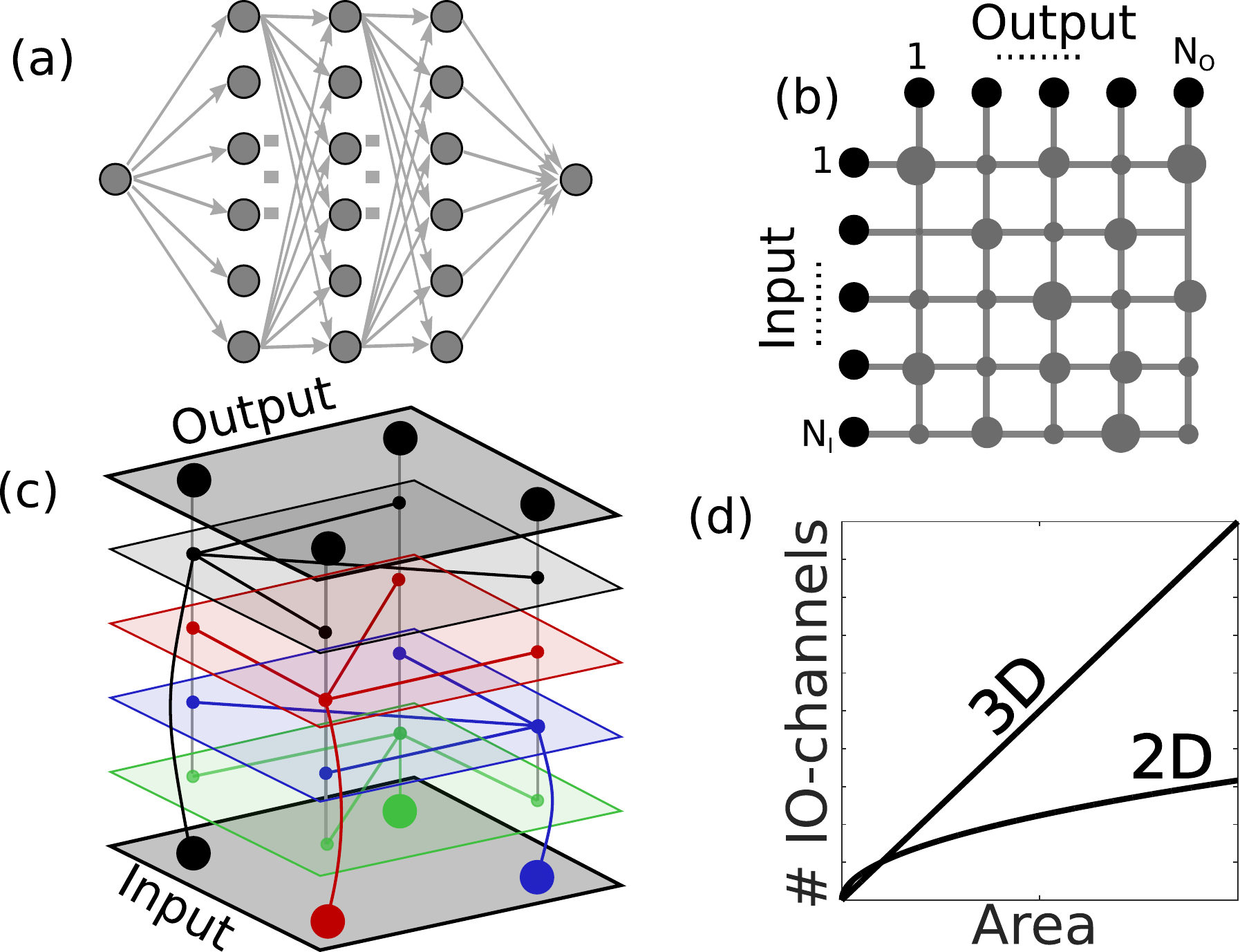}
	\caption{(a) Topology of a deep neural network.
		Links between layers of neurons correspond to large scale interconnects.
		(b) Crossbar arrays link in and output channels (IO-channels, black dots) in parallel in 2D; IO-channels are arranged along a line.
		(c) In three dimensions, IO-channels can be arranged in an array, while connections are implemented in the third dimensions.
		(d) The number of IO-channels of a parallel interconnect scales linearly with size in 3D.
		In 2D scalability is significantly worse.}
	\label{fig:3DScaling}
\end{figure}

A strategy to overcome many of the bottlenecks currently experienced in neural network computation is to realize integrated circuits adhering to a neural network's complex topology \cite{Appeltant2011,Nahmias2013,Shen2016,VanderSande2017,Brunner2019,Neckar2019}.
As schematically illustrated in Fig. \ref{fig:3DScaling}(a), a neural network  is formed by linking large numbers of nonlinear neurons, which often are grouped in layers.
It is particularly this intra-neuron interconnect which, despite recent progress \cite{Neckar2019}, still eludes a fully parallel and scalable hardware integration.
Most of today's integrated circuits are created via lithography, and are hence restricted mostly to 2D.
In cross-bar interconnects, see Fig. \ref{fig:3DScaling}(b), routing occurs via punctual contacts between two layers hosting input and output wires.
The $N_I$ input and $N_O$ output ports are arranged along a column or row, and hence their number scales with $N_I | N_O \propto \sqrt{A}$ for an area $A$.
This is the general behavior in 2D.

Three dimensional, additive manufacturing has significantly matured and allows complex structures with nanometric feature sizes \cite{Moughames2016, Deubel2004, Buckmann2012,Freymann2010}.
Crucially, the additional third dimension facilitates simple wiring topologies which are scalable, as schematically illustrated in Fig. \ref{fig:3DScaling}(c).
Input and output ports occupy a dedicated plane each (not rows or columns as in 2D), while the third dimension unlocks a circuit's volume for wiring: for each of the $N_I$ inputs, a dedicated plane hosts all connections to the $N_O$ outputs.
Even in such a simple routing scenario the system's scaling of area $N_I | N_O \propto A$ and height $H\propto N_I$ becomes linear.
The strong impact of 2D versus 3D integration on the scalability of a parallel interconnect is schematically illustrated in Fig. \ref{fig:3DScaling}(d).
Interestingly, the 3D routing strategy has been confirmed by evolution: the most reduced topological property of the human neocortex leverages the same effect.
Neurons are mostly located on its surface, while long range connections mostly traverse the volume.

However, 3D routing in electronics is challenging.
Lithographic fabrication requires of the order $N_I$ signaling layers, which makes such fabrication prohibitive for the kind of dimensionality demanded by neural networks.
Heat creation and heat dissipation from such a volumetric circuit's centre have additionally been identified as problematic \cite{Venkatadri2011}.
Disposing of this dissipated energy is a major bottleneck already for the mostly serial von Neumann processors \cite{Esmaeilzadeh2012}, and parallel interconnects for NN's require significantly more such layers and connections.
Photonics can overcome this challenge \cite{Lohmann1990,Miller2017}, which motivates the interest in photonic interconnects \cite{Lee1989,Shamir1989} and ultimately in photonic neural networks \cite{Farhat1985,Larger2012,Duport2012,Brunner2013a,Vandoorne2014,Shen2016,Pierangeli2019,Khoram2019} .

\section{3D interconnects of photonic waveguides}

\begin{figure}[t]
	\centering
	\includegraphics[width=\linewidth]{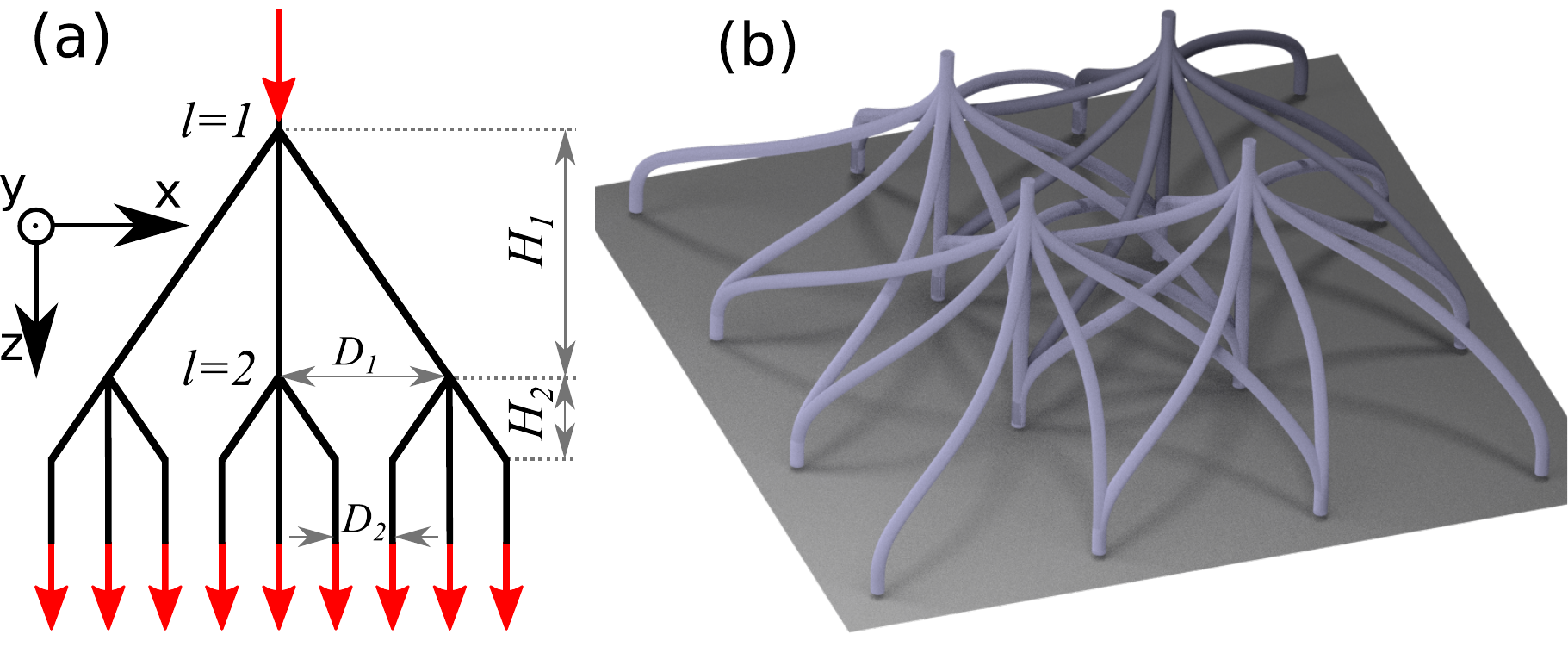}
	\caption{(a) Design principle of an optical coupler with a fractal geometry.
		Numerous layers of branching connections can be cascaded, and distances from one layer to the next scale with $\sqrt{b}$, where $b$ is the branching ratio.
		(b) 3D illustration of a small network hosting simple couplers.
		Chirality of the connections avoids the intersection of individual waveguides between the input and output ports.}
	\label{fig:fractal_tree}
\end{figure}

Low loss 3D printed photonic waveguides have been demonstrated at telecommunication wavelengths \cite{Lindenmann2012,Koos2013,Pyo2016,Nesic2019}.
Our waveguides were fabricated using a commercial 3D Direct-Laser writing system  from Nanoscribe GmbH (Germany).
A negative tone photoresist "Ip-Dip" dropped on a glass substrate (25x25x0.7~mm$^3$) was photopolymerized via two-photon absorption with a $\lambda=780~$nm femtosecond pulsed laser, focused by a 63X, (1.4 NA) objective lens.
After the writing process, the sample was immersed in a PGMEA (1-methoxy-2-propanol acetate) solution for 20 minutes to remove the unexposed photoresist.
Samples were written using the scanning mode based on a goniometric mirror, and the scanning speed on the sample's surface was kept constant at 10~mm$/\mathrm{s}$.
As optimization parameter we used the writing laser's power.
The diameter of individual waveguides is $d\approx 1.2~\mu$m, and they are spaced by $D_0=20~\mu$m.
Samples were structurally inspected using a scanning electron microscope (SEM, FEI 450W).
For optical characterization, we focused a 635~nm laser onto an input waveguide's top surface using a 50X microscope objective with NA = 0.8.
The mode field diameter of the focused beam is $\simeq2~\mu$m, hence larger than the input waveguide's diameter.
The emission at the couplers' output ports was collected by a 10X, NA=0.30 microscope objective and imaged onto a CMOS camera (iDS U3-3482LE, pixel size 2.2$~\mu$m) using a 100~mm achromatic lens, resulting in an optical magnification of 5.6.

\begin{figure}[tbp]
	\centering
	\includegraphics[width=0.9\linewidth]{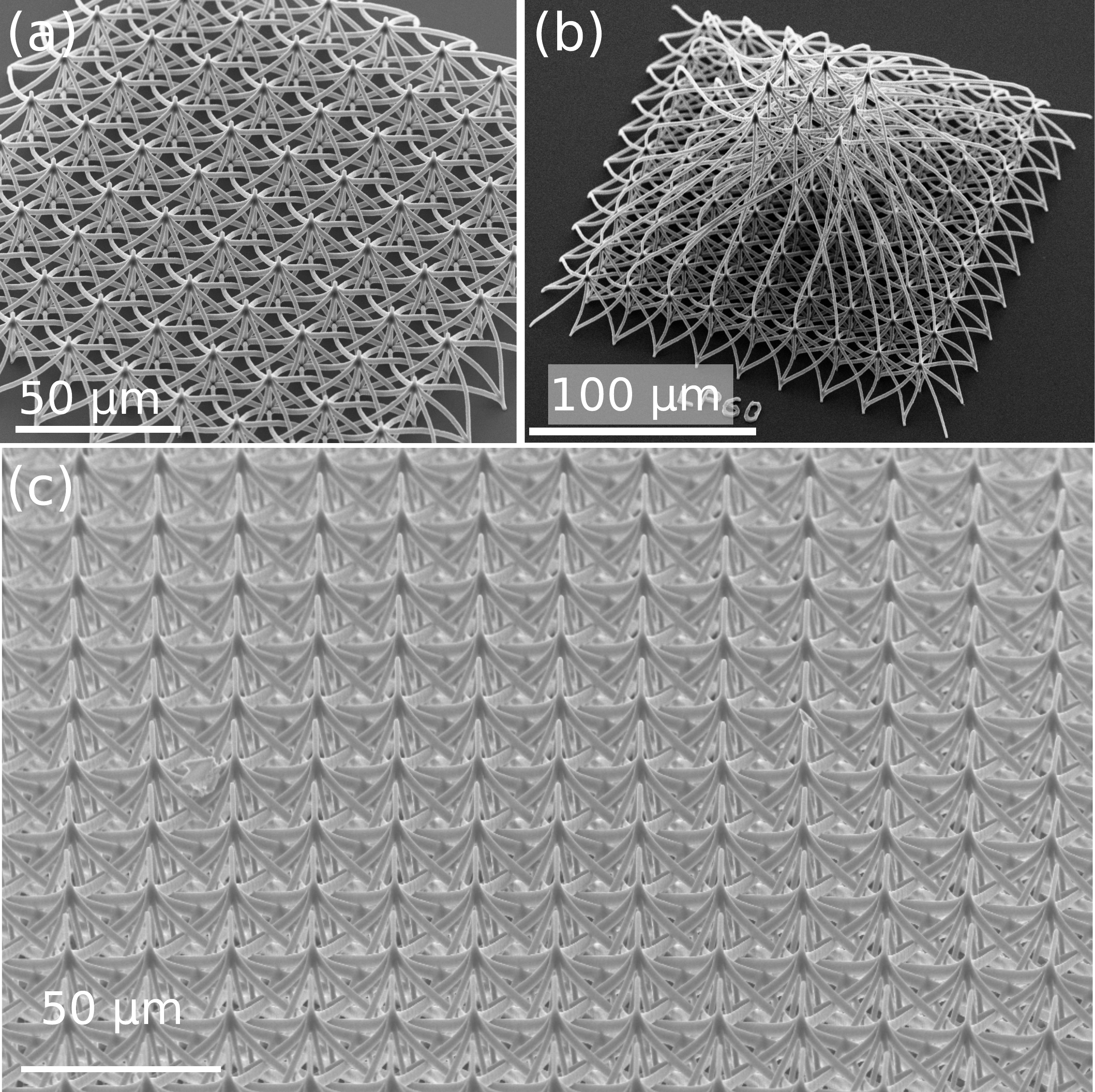}
	\caption{(a) SEM micrograph (15 kV, 40$^{\circ}$) of an array of $1\times9$ couplers hosting 9 elements.
		(b) SEM micrograph (10 kV, 40$^{\circ}$) image of 9 parallel $1\times81$ couplers featuring two bifurcation layers.
		(c) SEM micrograph (10 kV, 40$^{\circ}$) of a large scale array of $1\times81$ couplers, showing an area containing $15\times15$ units.}
	\label{fig:fractal_sem}
\end{figure}

\subsection{Fractal topology for fully connected layers}

Fully or densely connected layers are a principle topology in NNs \cite{LeCun2015,Jouppi2017}.
We adopt a routing strategy based on fractal (self-similar) branching, where each signal 'wire' splits into $b$ branches at the $l\in [1,\dots,L]$ branching points.
Figure \ref{fig:fractal_tree}(a) schematically illustrates such a fractal tree's 2D projection onto the $(x,z)$-plane for $b$=9 and $L$=2.
An input (top red arrow) is therefore distributed to $N_O = b^{L}$ output channels (bottom red arrows), here resulting in $N_O=81$.
Scaling of $N_O$ is therefore exponential in $L$, and $N_O=$6561 connections are created for each input channel for $b$=9 and $L$=4 branching layers only.
The tree's architecture is recursively defined according to the spacing between the $N_O$ output channels $D_L=D_0$ and height $H_L \propto D_0$.
The dimensions inside the bifurcation layers $l<L$ are $H_l = \sqrt{b}H_{l+1}$ and $D_l = \sqrt{b}D_{l+1}$.
Horizontal and vertical distances there scale identical, resulting in constant branching angles throughout the entire circuit.

This translation invariance aids the development of strategies to avoid the intersection of waveguides before layer $l=L$, where they merge into their respective outputs.
These details are illustrated for four neighbouring couplers with $b=9$ and $L=1$ in Fig. \ref{fig:fractal_tree}(b).
We incorporated chirality into the fractal couplers: the $b$ connections from a point in layer $l$ to layer $l+1$ have a negative curvature in the $(x,y)$-plane, which avoids intersections for vertical and horizontal connections.
Furthermore, avoiding intersections for diagonal links additionally requires curvatures in the $z$-direction.
% The parametrized curves and their coefficients are given in the supplementary material (citation).	
% \textcolor{red}{no suppl. matt. for the moment!}

Figure \ref{fig:fractal_sem}(a) shows an SEM image of a 3D fractal coupler array hosting $N_I=81$ input and $N_O=121$ outputs, each with $L=1$ and $b=9$.
We can see that chirality successfully avoids unintended intersections.
In Fig. \ref{fig:fractal_sem}(b) we show fractal trees for two bifurcations resulting in $1\times81$ coupling, with a circuit of $N_I=9$ inputs and $N_O=121$ outputs.
As for the single bifurcation layer 3D coupler, the two bifurcation layer couplers are mechanically sound, even though they feature waveguide sections with an aspect ration exceeding 50.
This excellent result motivated us to continue and integrate a full-scale interconnect with over 200 inputs, each of which are connected to 81 outputs, see Fig. \ref{fig:fractal_sem}(c).

\begin{figure}[tbp]
	\centering
	\includegraphics[width=\linewidth]{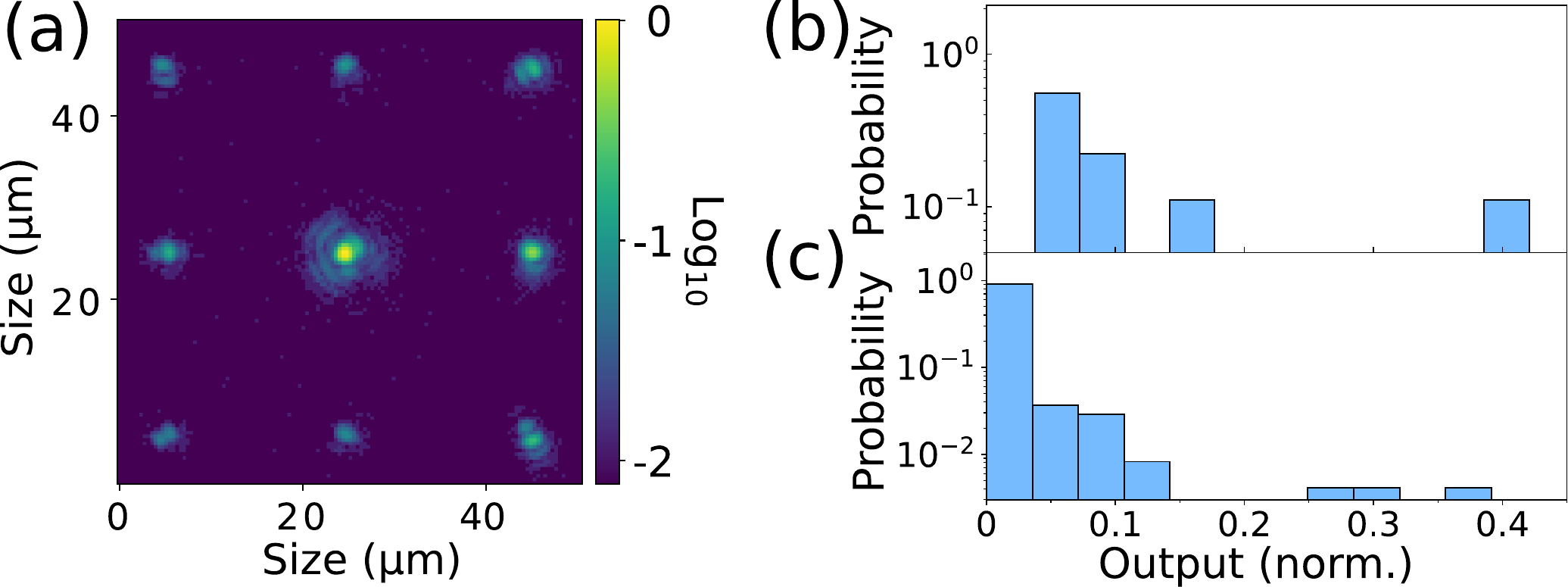}
	\caption{(a) Optical transmission through a single bifurcation layer $1\times9$ coupler, with intensity color-coded on a logarithmic scale.
		Histograms of the relative output intensity distribution for the $1\times9$ (b) and $1\times81$ (c) coupler.
		Statistical information obtained from three couplers, each.}
	\label{fig:OptCharact-Couplers}
\end{figure}

Figure~\ref{fig:OptCharact-Couplers} depicts the optical transmission through a $1\times9$ fractal coupler.
We used the camera images to characterize the optical losses and splitting ratios, where the injection spot focused onto the glass-substrate's top surface was used as reference.
The average optical losses for $1\times9$ couplers are 5.5~dB, which rise to 10.6~dB for $1\times81$ couplers.
Crucially, this includes optical injection losses $I$, propagation losses $P$ and losses induced at the coupling or bifurcation points $C$.
The fractal design principle allows us to determine each of these contributions.
As previously discussed, angles of the different bifurcation layers remain constant due to fractal design.
This results in identical bifurcation points for the entire topology, and hence we assume uniform coupling losses $C$.
Furthermore, we have characterized a $1\times9$ coupler with $\sqrt{b}=3$ times larger height $H_L$ and distance $D_L$.
This leaves us with three loss measurements (in dB), the standard $1\times9$ ($L_{1\times9}=I+C+P$), the three times larger $1\times9$ ($\tilde{L}_{1\times9}=I+C+3P$) and the $1\times81$ ($L_{1\times81}=I+2C+4P$) couplers, and we obtain  $I=2.71~$dB, $P=1.14~$dB and $C=1.67~$dB.

According to Fig. \ref{fig:OptCharact-Couplers}(a) some of the output ports' optical modes include second order Gauss-Laguerre contributions.
As our polymer waveguides are freestanding in air, they have an exceptionally high diffractive index contrast of $\Delta n\approx 0.5$.
According to the commonly employed approximation $M=0.5(\pi d \Delta n / \lambda)^2$  for the number of modes $M$ supported by a cylindrical waveguide, our waveguides support up to $M \approx 5$ optical modes.
However, early stage numerical simulations confirm that only the first and second optical mode are excited, which agrees with our experimental results.
We would like to point out that the high refractive index contrast allows (i) single mode waveguides with a diameter of 0.3$~\mu$m only, (ii) exceptionally narrow bending radii, and the combination of (i) and (ii) facilitates compact integrated photonic circuits.

We analysed three $1\times9$ and three $1\times81$ couplers with respect to the relative power distribution at their output ports, and statistical information is given in Fig.~\ref{fig:OptCharact-Couplers}(b,c).
For the $1\times9$ couplers we find that $(42 \pm 4 )\%$ of the total optical output power is provided by the central waveguide, with the remaining $\sim58\%$ quite evenly distributed among the off-center ports, see Fig. \ref{fig:OptCharact-Couplers}(b).
For $1\times81$ couplers, only $(33 \pm 6 )\%$ of the light is contained in the central waveguide, Fig. \ref{fig:OptCharact-Couplers}(c).
Interestingly, the $1\times81$'s ratio is not quite the square of the $1\times9$'s ratio, indicating that cascading our bifurcating waveguides cannot be fully approximated simply by linearly multiplying the coupling ratios of the individual components.
Higher order modes therefore appear to have an impact upon the splitting ratios.
Overall, the asymmetric splitting ratio is most likely caused by the geometry, and in particular by the branching angles of our waveguide couplers.

\begin{figure}[tbp]
	\centering
	\includegraphics[width=\linewidth]{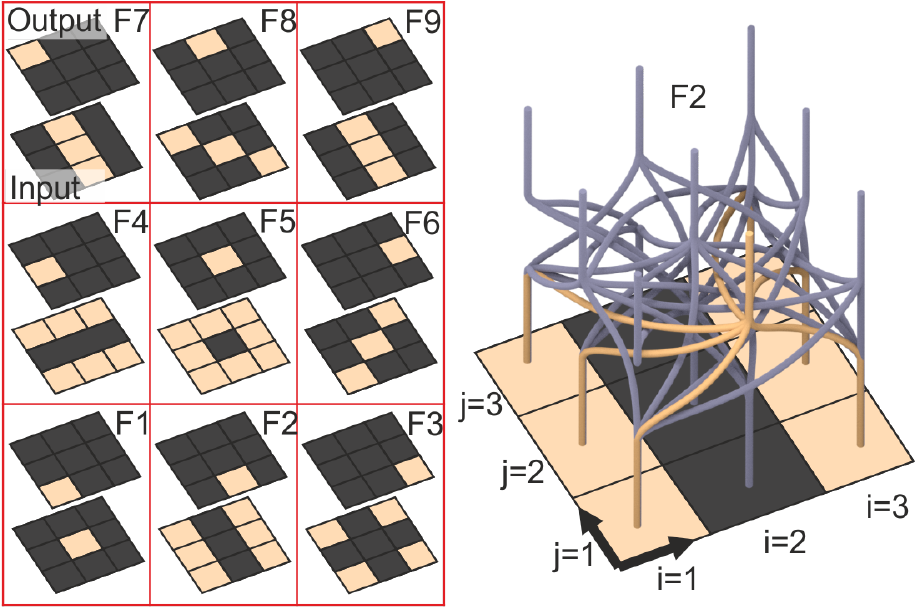}
	\caption{Schematic illustration of the input-output mapping of 9 Haar filters (F1-F9) with Kernel width and stride 3.
		A 3D-printed waveguide architecture realizing all 9 filters in parallel is shown on the left, with filter F2 highlighted in orange.
		The highlighted sub-stricture implements }
	\label{fig:haarfilter_scheme}
\end{figure}

\subsection{Haar filters}

The previously discussed, highly connected couplers, are typically required close to the output layer of deep neural networks.
However, their first layers often highlight structural aspects of input information by tailored, local connection topologies.
Examples are convolutional neural networks commonly employed in object recognition \cite{LeCun2015}.
Prominent convolution Kernels are so called Haar filters.
These feature 2D Boolean entries, and this simplification creates a sparse representation of information contained in images, which is a crucial operation for neural networks to be able to generalize to unseen test data \cite{NIPS2006_3130}.
We schematically illustrate in- and output properties of nine exemplary Haar filters (F1-F9) in Fig. \ref{fig:haarfilter_scheme}.
There, each filter Kernel's $3\times3$ Boolean weights (0: dark, 1:light) are illustrated as input, while each filter's dedicated output port is indicated as the output.

We developed a 3D routing topology, schematically illustrated on the right in Fig. \ref{fig:haarfilter_scheme}, to realize the 9 Haar filters.
Even in 3D this is challenging, which can be appreciated from the intricate network of connections.
Furthermore, the number of configurations scales factorial with the number of filters, and for the required 37 connections of the 9 filters there exist 362880 possibilities.
This already large numbers still ignores all geometrical aspects such as waveguide curvatures along the different dimensions.
In order to better illustrate the operating principle, we have highlighted the connection topology of filter F2 in orange.
For each filter, the input ports weighted by 1 are directly wired the the filter's output.
For incoherent injection into the $3\times3$ input waveguides, the intensity at the filter's output should therefore be proportional to the overlap between its Boolean weights and the input.

\begin{figure}[tbp]
	\centering
	\includegraphics[width=0.9\linewidth]{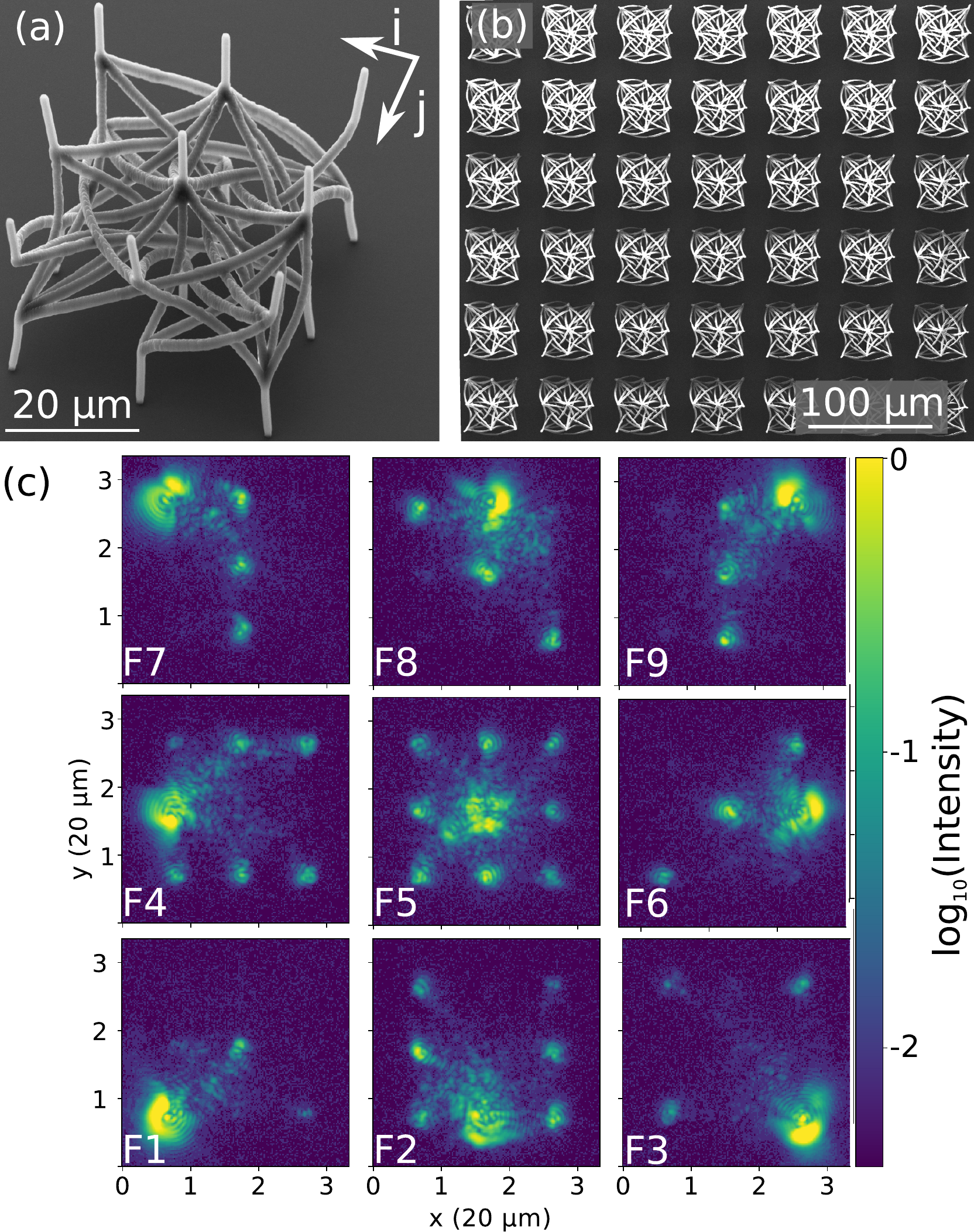}
	\caption{(a) SEM micrograph (10 kV, 40$^{\circ}$) of a single Haar-filter.
		(b) Full micrograph (5 kV, 0$^{\circ}$) of a large array hosting spatial-filtering for connecting layers of a convolutional neural network.
		(c) Optical characterization of the filter's connection topology, injection at the output port and recording the input ports emission.}
	\label{fig:haarfilter_sem_optics}
\end{figure}

In Fig. \ref{fig:haarfilter_sem_optics}(a) we show the SEM picture of the 3D printed spatial filtering interconnect realizing 9 Haar filters.
Waveguides feature smooth surfaces and the overall structure is stable.
However, one can identify a tendency that output waveguides with few connections start leaning outwards.
Figure \ref{fig:haarfilter_sem_optics}(b) shows a densely multiplexed array of Haar filter units.
Such an interconnect would implement the convolution of a $21\times21$-pixel input image simultaneously with filters F1-F9 fully in parallel.
As the individual filter units do not overlap in space the implemented convolution a convolution stride 3.

Figure \ref{fig:haarfilter_sem_optics}(c) depicts the optical characterization of the filters' connectivity using the same procedure as for the fractal optical couplers. 
The individual sub-panels correspond to the transmission through a different filter (F1 to F9) when injecting light into the output port. 
The optical characterization was therefore carried out in backward direction.
We opted for this procedure since output intensities of individual filters correspond to the filter's Kernel only in the backward direction.
In forward direction one would have to iteratively inject into the individual input ports and then sum the output intensities of the different injections; which is possible in principle yet less systematic.
Generally, we find an excellent agreement between the designed filter Kernels and the intensities recorded in the reverse propagation direction. 
The different loss mechanisms obtained for the fractal couplers are consistently reproduced for the Haar filters, with the peculiarity that each coupler exhibits distinct coupling losses $C$.
This, however, is to be expected; different filters rely on specific connection degrees, topologies as well as different branching angles.

There is some cross-talk from the optically injected port onto the image of the output plane.
One cause might be the smaller height of the overall 3D circuitry.
Light not collected and guided by the injected waveguide therefore illuminates a smaller area on the circuit's output plane, which in turn results in a higher intensity when imaged onto the camera.
The outwards-leaning input connections, see Fig. \ref{fig:haarfilter_sem_optics}(a), might additionally contribute.
The resulting non-orthogonal illumination of the waveguide's tip will most likely reduce the injection efficiency and therefore increase the cross-talk of uncollected light to the output plane.
For a fully integrated system this cross-talk will potentially be reduced significantly.
Inputs will in most cases be provided by optical fibers or waveguides arriving from an earlier stage of the optical system, for example when using a fiber bundle for collecting an input image.
This will also be true for the filter's output, which will be connected to some fiber or waveguide for further processing down stream.	

\section{Discussion and conclusion}

We successfully demonstrated complex and large scale 3D photonic interconnects.
Waveguides with a diameter of $\approx1.2~\mu$m were created by direct laser writing based on two photon polymerization.
Using this novel integration strategy we demonstrated intricate 3D routing topologies for large scale, highly connected as well as convoluting optical interconnects.
These example architectures were mostly oriented towards application in neural networks, where such interconnects can realize the large scale vector matrix products fully in parallel, with picosecond latency and potentially low energetic cost \cite{Peng2018}.
It is the first time that such complex and large scale integrated optical interconnects have been created in 3D.

As our concept scales linear in size it allows for novel routing topologies, which in turn will create new opportunities for integrated special purpose neural network chips.
Here, either complete implementations of neural networks, or the use of the photonic interconnects purely as neural network accelerators are a possibility \cite{Peng2018}.
However, there is a wider relevance for computing.
The end of Moore's and in particular Dennart's scaling is arguably induced by energy penalties of a processor's electronic signaling wires.
Photonic routing could prolong the scaling of classical electronic (or now: opto-electronic) von Neumann processors, and these ideas can be expanded to intra or inter chip connections.
Finally, non-computing related applications such as miniature remote sensing are further possibilities.
Ultimately we have demonstrated the first large scale 3D printed photonic circuit board.

The here reported findings are based on the first demonstrations of several, complex 3D photonic circuits, and performance as well as topologies offer significant potential for further improvements.
Beyond losses, it is in particular the asymmetric splitting ratios who deserve further attention, even though such imbalance can, by a certain degree, be compensated for by using phase-tunable topologies \cite{Miller2015}.
Couplers with an even splitting ratio (such as 1x4) promise potentially better homogeneity.

Most importantly, we have addressed the non-scalability of parallel and integrated interconnects for the first time.
In order to fully benefit from this new substrate, its functionalization is essential.
External control over a waveguide section's phase delay would enable unitary optical transformations on a scalable substrate \cite{Miller2015}.
An extension by active or nonlinear photonic elements will establish a new type of photonic device.
Crucially, small scale low bandwidth 3D printed polymer circuits are actively considered in electronics, for example for wareables \cite{Park2019}.

\section*{Funding Information}
The authors acknowledge the support of the Region Bourgogne Franche-Comt\'{e}.
This work was supported by the EUR EIPHI program (Contract No. ANR-17-EURE- 0002), by the Volkwagen Foundation (NeuroQNet I\&II), by the French Investissements d’Avenir program, project ISITE-BFC (contract ANR-15-IDEX-03) and partly by the french RENATECH network and its FEMTO-ST technological facility.
X.P. has received funding from the European Union’s Horizon 2020 research and innovation programme under the Marie Sklodowska-Curie grant agreement No. 713694 (MULTIPLY).

\section*{Acknowledgments}

The authors thank Marina Raschetti for technical support.

\section*{Disclosures}

M.T. works at the company Nanoscribe. The other authors declare no conflicts of interest.

% Bibliography
\bibliography{Bibliography}
	
\end{document}